\documentclass[11pt]{elsarticle}
\usepackage{amsmath,amssymb,graphicx,textcomp,multirow}
\usepackage{color}
\usepackage[section]{placeins}
\usepackage{graphicx}
\usepackage{multicol}

\usepackage{amssymb}

\journal{Applied Mathematics and Computation}
\begin{document}

\begin{frontmatter}



\title{Using Newton's method to model a spatial light distribution of a LED with attached secondary optics}


\author[david]{David Kaljun}
\author[david]{Jo\v ze Petri\v si\v c}
\author[david,imfm]{Janez \v{Z}erovnik}

\address[david]{FS, University of Ljubljana, A\v{s}ker\v{c}eva 6, 1000 Ljubljana, Slovenia\\
      david.kaljun@fs.uni-lj.si}
\address[imfm]{Institute of Mathematics, Physics and Mechanics, Jadranska 19, Ljubljana, Slovenia\\
      janez.zerovnik@fs.uni-lj.si}

\begin{abstract}
In design of optical systems based on LED (Light emitting diode) technology, a crucial task is to handle the unstructured data describing properties of optical elements in standard formats. This leads to the problem of data fitting within an appropriate model. Newton’s method is used as an upgrade of previously developed most promising discrete optimization heuristics showing   improvement of both performance and quality of solutions. Experiment also indicates that a combination of an algorithm that finds promising initial solutions as a preprocessor to Newton’s method may be a winning idea, at least on some datasets of instances. 
\end{abstract}


\end{frontmatter}

\textbf{Highlights:}
\begin{itemize}
  \item Model for data fitting of LED photometry with the evaluation function is presented.
  \item The effects of a numerical method in conjunction with heuristics are studied.
  \item Algorithms are developed with the use of C++ programing language.
  \item The success of developed algorithms is tested on real and artificial datasets.
  \item The results are statistically evaluated. 
  \item The numerical Newton’s method prevails on both datasets, and provides substantial runtime shortening.
\end{itemize}


\section{Introduction}
\label{intro}
The LED (Light emitting diode) industry has been evolving rapidly in the past  several years. 
The fast pace of research and development in the field     had  some expected  impact.
One of  the results is a  massive use and implementation od LED elements in all kind of luminaires. While some of these luminaires are designed for ambient illumination the majority are technical luminaires that have to conform not only to electrical and mechanical safety regulations but also to regulations that define and restrict the photometry of a certain luminaire. This means that the photometry of a luminaire has to be defined prior to the production. In order to   do that efficiently and with minimal errors the design engineer must virtually test the luminaires performance. Tools that can be used (OpticsWorks, LigthTools, TracePRO)
 \cite{optis,lighttools,tracepro}
 do exist and they offer a vast repository of sub-modules to develop and design custom lenses,  reflectors, light guides, etc. These universal tools however do not completely exploit the luminaire design possibilities that were introduced by the transition from conventional light source technologies to LED. One of the possibilities which is also the main goal of a bigger study that incorporates the research presented here is to have an expert or intelligent system  which would be capable of suggesting a secondary lens combination that would result in a user defined end photometry. In other words, the system would take some stock secondary LED lenses from different manufactures, place them on a defined LED array and search for the optimal combination of the lenses so that the resulting photometry would be as close as possible to the user defined one.
%
%
 
 The method could enable the luminaire designer to custom design the light engine to a specific area of illumination, while keeping the mechanical and electrical parts of a luminaire untouched. This would in turn provide a customer with a tailored solution that would guarantee a maximum efficiency, lower prices, fewer light pollution and the possibility to individualize the illumination effect while maintaining a consistent visual appearance of the luminaries. 
There are several optimization tasks related to development of the above idea. 
Here we focus on the approximation of spatial light distribution with a moderate number of suitable  basis functions \cite{OpticsExpress,moreno_sun}.   
The problem that is defined formally in the next section is motivated by the following.
The data describing the properties of the lenses and/or of the desired light distribution is nowadays usually given in some standard format files that correspond to the measured (or desired) values at a   number of points in space. 
This results in relatively large data files of unstructured data.
Clearly, if the data can be well enough approximated c.f. as a linear combination of certain basis functions, this may   enable faster 
computations using less computer storage. Indeed, for some special cases including LED lenses with symmetric light distribution, it is possible to find reasonably good approximations fast (8 minutes runtime on a Intel Core I7-4790K CPU @ 4 Ghz, the code is written in C++ and is not fully optimized). 
Sufficiently good approximation here means 2-5\% RMS error (to be defined later) for target light distribution,
taking into account expected noise in measurement using   current technology.
Recent  experiments showed that  
sufficiently good approximations can be  obtained by some  basic optimization  algorithms, including local search algorithms and genetic algorithms \cite{osijek,informatica,chapter}.    
However, when using predefined lenses to  design a luminaire  that closely approximates a  desired light distribution,
it may be essential that the approximation error is much lower. 

The same task can also be seen as solving a problem of data compression, replacing a long unstructured data file with 
a much shorter one, in this case a sequence of parameters. 
It makes sense to aim at 0\% approximation when considering the data compression task. 
 
As the functions to be approximated are smooth,  it is natural to try to improve the basic discrete optimization methods with 
continuous optimization techniques, c.f.  Newton's method \cite{newton}. 
Here we consider Newton's method both as a standalone (restarted) algorithm and as a post-processor of other   algorithms.
The   datasets  used for testing and analysis 
are a selection of real lenses as used in previous studies and an artificial dataset that is large enough for statistical analysis. 
The artificial dataset is also generated in a way which  assures   that 0\% approximation is possible.
Note that  we have no guarantee that the realistic lenses may be approximated within our model with arbitrary low RMS  error. 
The rest of the paper is organized as follows. In section two we discus the problem and present the mathematical model, section three is all about the algorithms and Newton method implementation, section four presents the datasets used in the experiment, section five provides the experiment set-up, section six unveils the results section seven wraps everything up in the conclusion. Appendix provides all equations needed for the Newton method.

\section{The model}

	The method mentioned above seems natural and straightforward, but 
at a closer look, we observe some fundamental problems related to realization of the main idea.
Namely, both the spatial light distribution of LED lenses and the desired illumination are given in the standard data formats, that are just a long 
unstructured lists of data. In particular when the aim is to construct a lighting system that provides the desired illumination of the environment, 
it is necessary or at least very convenient to have the data in some more structured format. It is known that the spatial light distribution of 
some LED lenses can be approximated by a sum of a small number of certain basis functions \cite{OpticsExpress}.
Provided the approximation is sufficiently good, it may be possible to provide designs combining several lenses with controlled error rate.

This naturally opens several  research avenues. For example, it is important to have error free or at least very good approximations of the basic lenses,
and to have methods that are stable in the sense that they are not too sensitive to the noise in the presentation of basic elements.

\begin{figure}[h]
\centering
\includegraphics[width=8cm]{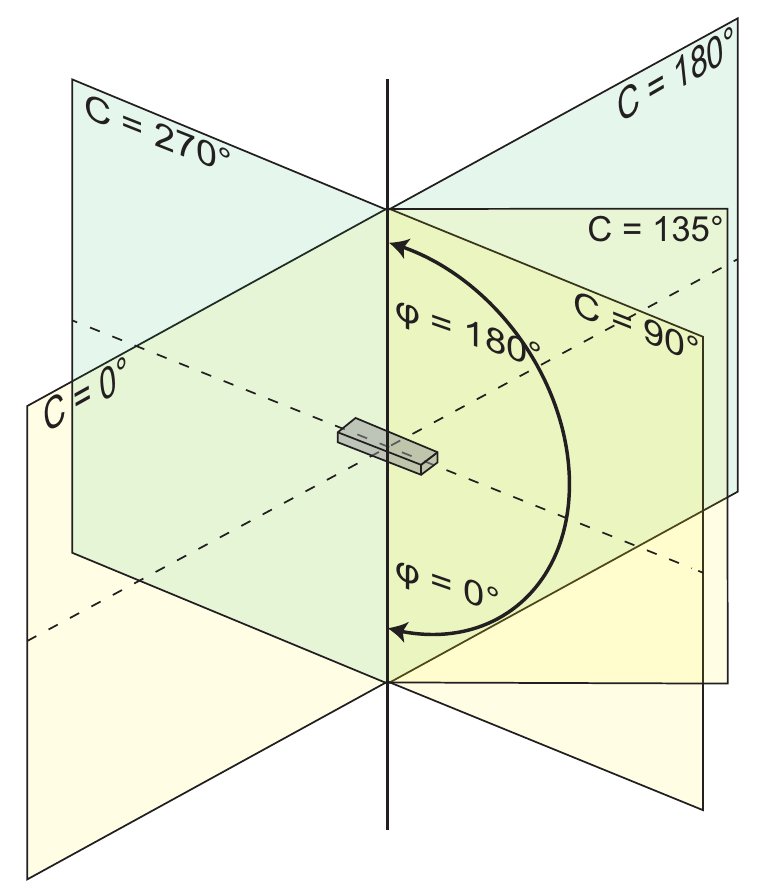}
\caption{C-planes according to standard. C-planes angles : 0° - 360° | $\varPhi$ angles : 0° to 90°}
\label{Cplanes}
\end{figure}

Here we focus on the first above mentioned task, approximation of the unstructured spatial light distribution data. 
We search  for an  approximation of   the   Luminous intensity $ I\left( \varPhi;  \textbf{a},\textbf{b},\textbf{c}\right) $ at the polar angle of $\varPhi$
in the form 
{\small \begin{equation}
I\left( \varPhi;\textbf{a},\textbf{b},\textbf{c}\right)=I_{max}\sum_{k=1}^{K}\ a_{k}\cos^{c_{k}}(\varPhi-b_{k})
 \label{3} 
\end{equation}}
where 
 $K$ is the number of functions to sum and $a_k$, $b_k$, $c_k$ are the function coefficients that we search for. 
For brevity, coefficients are written as vectors 
$\textbf{a} = (a_1,a_2, \dots, a_K)$,
$\textbf{b}= (b_1,b_2, \dots, b_K)$,  and 
$\textbf{c}= (c_1,c_2, \dots, c_K)$.
The interval range of the coefficients is: $a = [0, 1]$, $b = [0,90 ]$ and $c = [0,  100 ]$, more accurately the discrete values are : $a_\star \in [0,0.001, 0.002,\dots,1]$, 
$b_\star \in [-90,-89.9,-89.8,\dots,90]$, and 
$c_\star \in [0,1,2,\dots,100]$.
Here we need to note two restrictions on the model. First restriction emerges from the LEDs physical design. The LED can not emit any light to the back side which is the upper hemisphere in our case. That is why all intermediate values that are calculated at the combined angle $(\varPhi-b_{k})$ greater than 90\textdegree  ~equal 0. 
The second restriction deals with the slightly unusual description of the light distribution in standard files such as Elumdat 
(file extension .ldt) \cite{sist_en} and Iesna (.ies) \cite{IESNA}. These files present measured candela values per angle $○\varPhi$ on so called C planes which can be observed on Figure \ref{Cplanes}. One C plane is actually only one half of the corresponding cross-section and does not describe the other half. But from a physical point of view we need to consider the impact from the other half of the cross-section.
Because all  the lenses used  here  are symmetric, we can simplify the calculation of the intermediate values and incorporate the impact of the other half by mirroring (multiplying by -1) all values that are calculated with the combined angle $(\varPhi-b_{k})$ less than 0°. Note however that this only works with symmetrical distributions,
and should be reconsidered carefully  when  the method is to be applied  to asymmetrical distributions.

The goodness of  fit is defined  as  the root mean square error ($RMS$), formally defined by the expression: 
{\scriptsize \begin{equation}
 RMS \left(\textbf{a},\textbf{b},\textbf{c}\right) = \sqrt{\frac{1}{N}\sum_{i=1}^N \left[ I_m(\varPhi_i) - I(\varPhi_i,\textbf{a},\textbf{b},\textbf{c})\right] ^{2}}  
\label{RMS}
\end{equation}}
where $RMS$ represents the error of the approximation.Later in tables we provide the relative RMS error (RMSp) defined with  equation \ref{RMSp}. $N$ is the number of measured points in the input data, 
$ I_m(\varPhi_i)$ the measured Luminous intensity value at the polar angle $\varPhi$ from the input data, 
and $I(\varPhi_i,\textbf{a},\textbf{b},\textbf{c})$ the calculated Luminous intensity value at the given polar angle $\varPhi$. 

{\scriptsize \begin{equation}
 RMS_{p} \left(\textbf{a},\textbf{b},\textbf{c}\right) = \frac{ 100 * N* RMS\left(\textbf{a},\textbf{b},\textbf{c}\right)}{\sum_{i=1}^N \left[ I_m(\varPhi_i)\right]} \left[\%\right]
\label{RMSp}
\end{equation}}

{\bf Remark.}
The model was  successfully applied to  LED's with attached secondary optics and symmetric light distribution \cite{OpticsExpress}
showing that sufficiently good approximations
(RMS error below 5\%) 
 can be obtained using a sum of only  three functions  ($K=3$). 
%
%
%
%
Approximation of spatial light distribution of a LED with uniform distribution and without a secondary lens  using this type of functions was first  proposed in \cite{moreno_sun}.
 The model was slightly modified in  \cite{OpticsExpress}
 where a new normalizing parameter was introduced, and consequently, all other parameters will have values in fixed intervals known in advance. 
It should be noted that the modified model is equivalent to the original, only the number of parameters  and their meaning differ.
It may be interesting to note  that due to symmetries of the examples, $K=3$ is sufficient  for both applications \cite{OpticsExpress,moreno_sun}.
In general case, we expect that $K>3$  functions will be needed for sufficiently good approximations, and in view of optimization of the design of a luminaire it
is interesting to have an idea how large the parameter $K$ can grow to assure that the light distribution  fits the desired (and/or standard) sufficiently well.
We do not address this question here.

\medskip
When applying the model to the data compression problem, the target RMS error is 0\%.
Therefore, we aim to improve the approximation results that were obtained previously \cite{osijek,informatica}
and  restrict   attention to symmetric light distributions. 
Also, we fix $K=3$ functions in the model.
Besides the dataset of 14 realistic lenses that was used in some previous studies, here
we also generate an artificial dataset in which a sample  is simply a sum of three basis functions with randomly chosen  parameters.
This assures that zero error approximation is possible for the instances of the artificial dataset.

We are interested first in minimizing the approximation error, and second, in computational time of the methods. 
In the next section we briefly outline the algorithms we use in the experiments.

\section{The algorithms}

In previous work  \cite{osijek,informatica},   the model described above
was applied  in conjunction with several custom build algorithms that are based on local search heuristics    and some  meta-heuristics. 
The algorithms implemented  include 
 a steepest descend algorithm, two iterative improvement algorithms with different neighbourhoods and two genetic algorithms,
 a standard one and a hybrid one in which the best individuals of  
 every generation  are optimized with the iterative improvement algorithm.
For more detailed  description of the algorithms  we refer to  \cite{osijek,informatica}.
The results of the experiments  showed that 
all of  the  algorithms  applied are capable of providing satisfactory results on all tested instances, 
and differed mainly  in computational time needed. 
The average RMS values obtained  on real lenses were around $ RMS = 2\%$.
Hence,  the results mentioned proved that the model is accurate and that sufficiently good  approximations can be found
with a variety of algorithms  for sufficiently good description of lenses.

However, recall that the model can also be used for data compression task.   
Zero or very low RMS error is also  essential in the foreseen application, in which 
the pre manufactured lenses are to be combined into a more complex luminaire with prescribed light distribution.

In  the  model  we use   a  sum of functions that  are smooth
and hence   the first and second derivatives can be calculated
allowing application of   continuous optimization    methods  in addition  to  the
general  discrete optimization   meta-heuristics  that were  used before.
 We have chosen to use the Newton (also known as the Newton\textendash Raphson) iterative method   \cite{newton}  to find the solution that we seek.
 It is well known that convergence of the Newton method largely depends on the initial solution.
Therefore  we have applied  the method in two  ways. 
First, we use the Newton method as  an optimizer which will pinpoint the local minimum of the solutions found by heuristic algorithms. 
In a sense this implementation of the Newton method will be an extension of the discrete optimization algorithm, used  to finalize the search to
end in a local minimum. 
(Note that the local minima may be missed by the discrete optimization algorithms due to predefined length of the discrete  moves.)
Second, we  use the Newton method  as a standalone algorithm that will on initialization generate a number of random
(initial)
 solutions that are uniformly scattered over the whole search space and then it will use the Newton method on 
a number of  the  
 best initial solutions to find the local minimums. 
Of course,
 for both implementations to be comparable  the iteration count
has to be controlled  so that the overall maximum amount of
computation time 
will be roughly the same.

 {\bf Preprocessor multi-start IF.} The \textbf{ multi-start iterative improvement with fixed neighbourhood (IF)} algorithm \cite{informatica,chapter} first initializes several initial solutions. The initial solutions are randomly chosen from the whole search space. Each of the initial solutions is then optimized using the following steps. In the beginning the search step values (a step for a numerical differentiation) $da = 0.01$,  $db = 1$, and  $dc =  \frac{I_{max}}{10}$ are initialized,  
giving the 512 neighbours of the initial solution: 
($a_1\pm da, b_1\pm db, c_1\pm dc$, $a_2\pm da, b_2\pm db, c_2\pm dc$, $a_3\pm da, b_3\pm db, c_3\pm dc$).
Then the algorithm randomly chooses a neighbour, and immediately moves to the neighbour if its RMS value is better than the current RMS value.
 If no better neighbour is found after 1000 trials, it is assumed that no better neighbour exists. In this case the algorithm morphs the neighbourhood by changing the step according to the  formula $d_{i+1}=d_{i}+d_{0}$.
More precisely,  $da_{i+1}=da_{i}+da_{0}$ where $da_{0}$ is the initial step value. Analogously for $db$ and $dc$.

 This is repeated until $i=10$. If there still is no better solution, the initial step value is multiplied by $0.9$ and the search resumes from the current solution with a finer initial step. The algorithm stops when the number of generated solutions reaches $T_{max}$.

{\bf Newton method.}
Newton’s method \cite{newton,newton1,newton2} is a well-known numerical optimization method that can provide very good results under certain assumptions on the evaluation function and on the initial solution. Newton’s method indirectly minimizes the evaluation function by looking for a solution of a system of nonlinear equations (first derivatives of the evaluation function).  Newton’s method solves the system of nonlinear equations iteratively by approximating it with a system of linear equations in each step which produce the delta vector. The delta vector is a part of the iterative sheme $\textbf{x}_{k}^{i+1}=\textbf{x}_{k}^{i}-\textbf{d}_{k}^{i}$. Newton method converges when the delta vector vanishes, $d=0$. At this point the evaluation coefficients found are the local minimum. Details are given in the Appendix. An obvious assumption is that the evaluation function has to be a continuous non-linear function for which first and second order derivatives are defined. The initial solution has to be close enough to a local or global optimum, for Newton’s method to converge. Hence the method is very sensitive to the choice of the initial solution. 

\section{The datasets}

The experimental study uses two batches of instances, a dataset of 14 instances that correspond to real LED lenses, and a 
dataset of artificial instances generated for purpose of this experiment.
The artificial lenses are used to obtain more conclusive result on the statistical test, 
because a sample of 14 is rather small and may provide
statistically insignificant  results. 
The real lenses on the other hand show  that the algorithms are useful in real life scenarios.    

\medskip\noindent 
{\bf Real lenses.} 
We have chosen 14 different symmetrical lenses which are meant to be used with a CREE XT-E series LED, from one of the world's leading lens manufacturer LEDIL from Finland. We acquired the photometric data from LEDIL's on-line catalogue \cite{LEDIL} The data was provided in .ies format, which we then converted to a vector list which is more suitable to use in our algorithms. LEDIL measured the individual lenses with a 1° polar precision on four C panels. This means that from every .ies file we extracted 720 vectors. As the lenses are symmetric we only needed one C panel, and because we are only working on the lower half of the sphere (DLOR) we end up with 91 vectors (counting the 0° vector) on which we approximate the model.

\medskip\noindent 
{\bf Artificial lenses.}
In the  dataset of  100 artificial examples,   each element in the dataset was generated as follows. 

A value from an  interval
was generated using uniform random distribution.
 (Intervals are $[0,1]$, $[0,90]$, or  $[0,10]$, depending on the parameter.
More precisely, the random generator chose one of the values from the finite sets:
$a_1,a_2, a_3 \in \{0, 0.001, 0.002, \dots,0.999, 1\}$, $ b_1, b_2, b_3 \in \{0,0.01,0.02,$ $ \dots,$ $90\}$ and 
$c_1,c_2, c_3  \in  \{0,0.1,0.2,$ $\dots,$ $10\}$.)
Then the function values or candela values were computed for each polar angle $\varPhi  \in  \{0,1,2, \dots, 89\}$. The candela values for  polar angles $\varPhi  \in  \{90,91,92,\dots,180\}$ were set to 0. The data was then encoded into a .ies file structure, giving a datafile of the same format as the real lenses have. Note that the data generated assure that in each case zero RMS error approximation is possible within our model.
Second, the dataset of 100 samples is sufficiently large for meaningful statistical analysis of experimental results.

\section{The experiment setup }

Before we go ahead and explain the experiment set-up, let us first remember the evaluation function that is the basis of the Newton method \cite{newton}. We already showed that the goodness of   fit is measured with the $RMS$ value which is calculated from (\ref{RMS}). From that we can define the evaluation function as:

\begin{scriptsize}
\begin{equation}
 E \left(\textbf{a},\textbf{b},\textbf{c}\right) 
= \frac{1}{N}\sum_{i=1}^N \left[ I_{max} ( a_{1}\cos^{c_{1}}(\varPhi-b_{1}) +a_{2}\cos^{c_{2}}(\varPhi-b_{2}) +a_{3}\cos^{c_{3}}(\varPhi-b_{3}) ) -  I_m(\varPhi_i)   \right] ^{2}  
\label{eval}
\end{equation}
\end{scriptsize}

\noindent where $E$ represents the error to be minimized, $N$ the number of measured points in the input data,   $I_{max}$ the maximum candela value,
$I_m(\varPhi_i)$ the measured Luminous intensity value at the polar angle $\varPhi$ from the input data, and $I(\varPhi_i)$ the calculated Luminous intensity value at the given polar angle $\varPhi$. 

The experiment was set-up to provide data from different algorithms.  
 This in turn enables an objective comparison and a statistical test to determine the best algorithm. Recall that we implemented the Newton method in two distinct ways. The first implementation uses the multi-start version of iterative improvement (\textbf{IF}) to find a good approximation which is then optimized via the Newton method. The second implementation uses the random generator to generate initial solutions   of which 100 best are optimized with the Newton method. Table below
shows different algorithms that were prepared for the experiment.

\begin{table}[hhtb]
\caption {Experiment algorithms.}

\centering
  \begin{tabular}{|c|c|c|c|c|c|}
    \hline
    \textbf{Config.}&	\textbf{Algorithm}& \textbf{Multi-start}& \textbf{IF steps}\\ \hline    
    \multicolumn{4}{|c|}{\textbf{Short runs 1 million}} \\ \hline   
    \textbf{1}&		\textbf{S-Newton}&	1000000&	NA\\ \hline
    \textbf{3}&		\textbf{IF10}&	10&	100000\\ \hline
    \textbf{4}&		\textbf{IF20}&	20&	50000\\ \hline
    \textbf{5}&		\textbf{IF50}&	50&	20000\\ \hline
    \textbf{6}&		\textbf{IF100}&	100&	10000\\ \hline
    \multicolumn{4}{|c|}{\textbf{Long runs 4 million}} \\ \hline
    \textbf{2}&		\textbf{L-Newton}&	4000000&	NA\\ \hline 
    \textbf{7}&		\textbf{IF40}&	40&	100000\\ \hline
    \textbf{8}&		\textbf{IF80}&	80&	50000\\ \hline
    \textbf{9}&		\textbf{IF200}&	200&	20000\\ \hline
    \textbf{10}&	\textbf{IF400}&	400&	10000\\ \hline
        
  \end{tabular}
\label{eksper}       

\end{table}
\medskip\noindent  {\bf  Time.}
We ran the algorithms in two different lengths. The short run evaluates approximately one million possible solutions per instance (lens) in just under 45s, and the long run approximately
 four million possible solutions per instance in about 3 minutes on a Core I7 - 4790K CPU. 
Newton’s method took in average 3 to 4 iterations to converge which means that the time it took to run the Newton’s method is negligible in contrast to the time the whole algorithm run. Expressed in seconds, the Newton’s method took approximately $2* 10^{-3}s$, opposed to minutes CPU for the heuristics. In addition to the different time/iteration spans we ran the algorithms on two instance sets.

\medskip\noindent  {\bf  Datasets.}
Recall the two datasets of instances explained above, the dataset of 14 real lenses and a the dataset of 100 randomly generated artificial instances.

\medskip\noindent {\bf Algorithms.}
We apply  the Newton  method both as a standalone algorithm (restarted on randomly generated initial solutions)
and as a final step after discrete local search algorithm (\textbf{IF}) outlined  above.
There are several algorithms that vary in the number of multi-starts (or, equivalently in the length of each  local search).
Depending on the length (short run, long run) and the number of restarts we denote the algorithms 
by {\bf IF10, IF20, IF50, IF100} and 
by {\bf IF40, IF80, IF200, IF400}.
The versions without local search are denoted by {\bf S-Newton} and    {\bf L-Newton} for short and long runs, respectively. 
See Table \ref{eksper}.

\section{Experimental results}

We will begin the section with a comparison of the raw  experimental data followed by the performance (quality of results) ranking and finish with the results of the  Wilcoxon Signed rank test.

\begin{table}[h]
\caption {Artificial lenses statistical data in RMSp for short and long runs. Best two results are emphasized.}

\centering
  \begin{tabular}{|c|c|c|c|c|}
    \hline
    \textbf{Algorithm}&	\textbf{Mean}& \textbf{Std. dev.}& \textbf{Min.}& \textbf{Max.}\\ \hline    
    \multicolumn{5}{|c|}{\textbf{Short runs 1 million}} \\ \hline
   {\bf S-Newton} & {\bf 1.3797E-04} & 8.5102E-05 & {\bf 3.5054E-05} & 3.9148E-04 \\ \hline
   {\bf IF 10}    & 2.3393E+01       & 3.0447E+01 & 5.8079E-05       & 2.4287E+02 \\ \hline
   {\bf IF 20}    & 1.1211E+01       & 1.6382E+01 & 4.9583E-05       & 1.3287E+02 \\ \hline
   {\bf IF 50}    & 6.3340E+00       & 9.2152E+00 & 4.1291E-05       & 6.8720E+01 \\ \hline
   {\bf IF 100}   & {\bf 7.2277E-01} & 2.7120E+00 & {\bf 3.4934E-05} & 1.9051E+01 \\ \hline
    \multicolumn{5}{|c|}{\textbf{Long runs 4 million}} \\ \hline
    {\bf L-Newton} & {\bf 1.3797E-04} & 8.5102E-05 & {\bf 3.5054E-05} & 3.9148E-04 \\ \hline
    {\bf IF 40}    & 1.2085E+01       & 2.5447E+01 & 4.8905E-05       & 2.4287E+02 \\ \hline
    {\bf IF 80}    & 1.7490E+00       & 4.2982E+00 & {\bf 3.5054E-05} & 1.9504E+01 \\ \hline
    {\bf IF 200}   & {\bf 4.0423E-01} & 1.9606E+00 & {\bf 3.5054E-05} & 1.4237E+01 \\ \hline
    {\bf FI 400}   & 6.1717E-01       & 2.4068E+00 & {\bf 3.4934E-05} & 1.8938E+01 \\ \hline 
  \end{tabular}
\label{stat01}       

\end{table}

\begin{figure}[h]
\centering
\includegraphics[width=15cm]{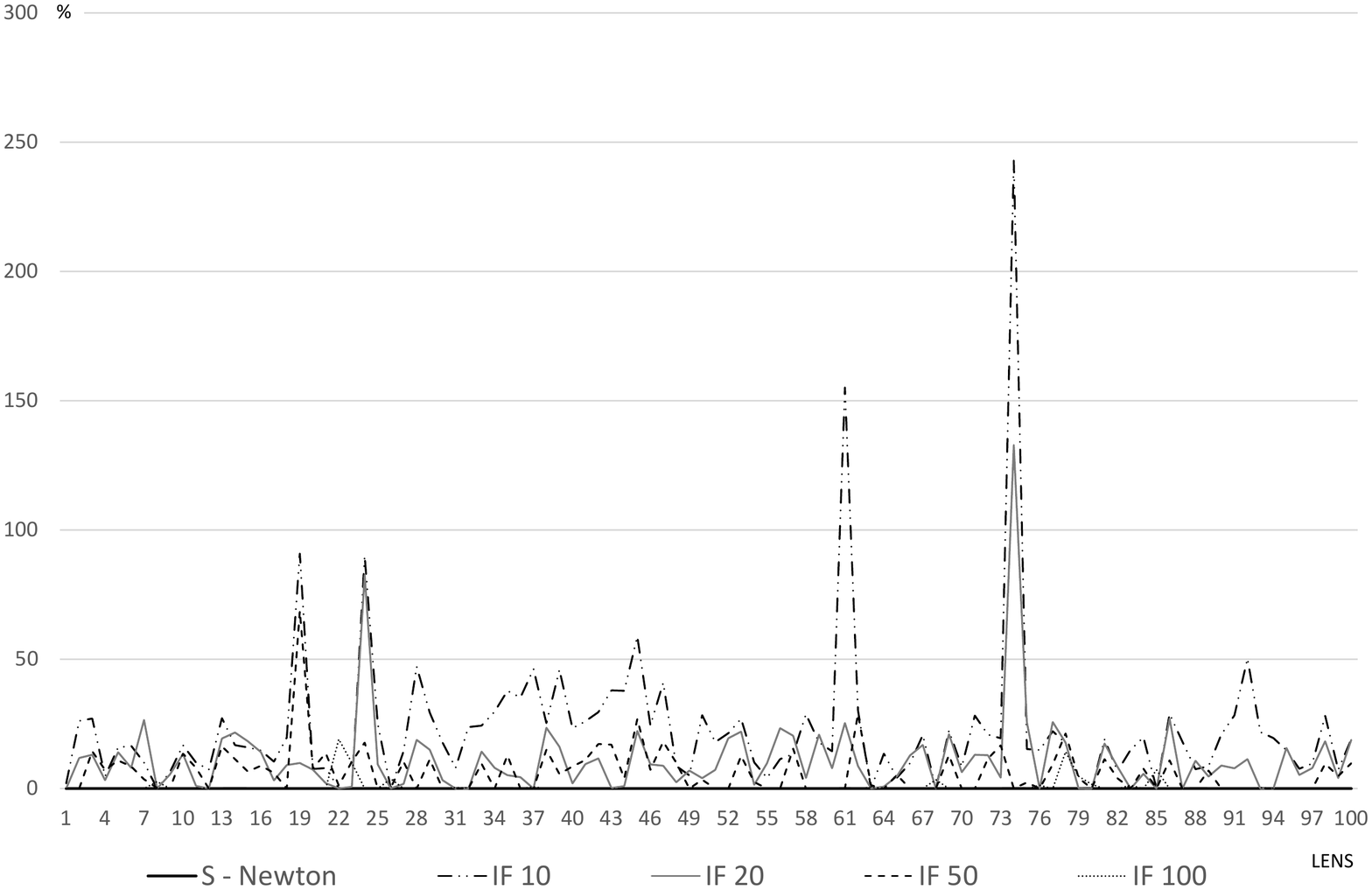}
\caption{Best found solution on a short run. Artificial lenses per algorithm.}
\label{shortart}
\end{figure}

\begin{figure}[h]
\centering
\includegraphics[width=15cm]{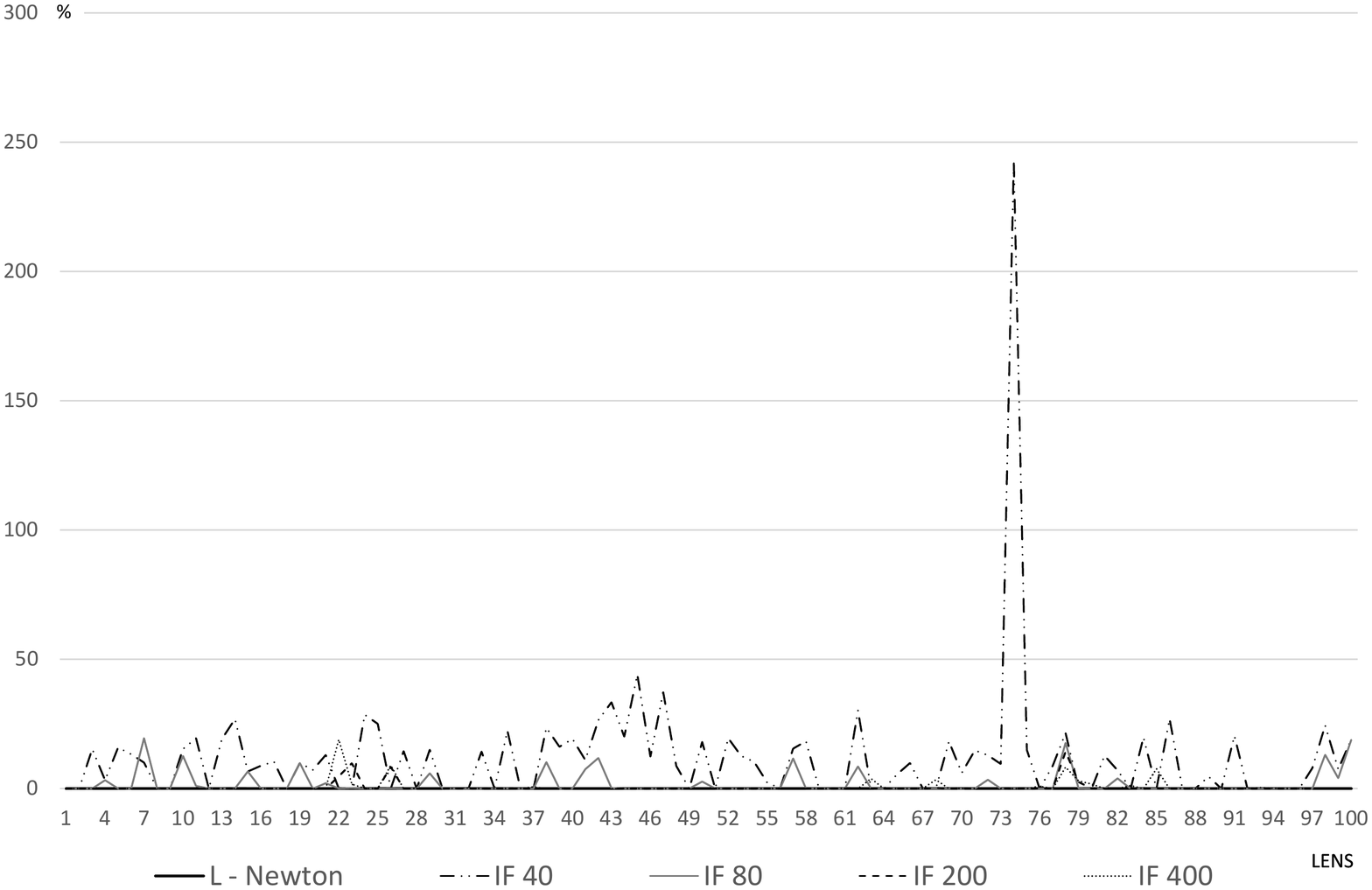}
\caption{Best found solution on a long run. Artificial lenses per algorithm.}
\label{longart}
\end{figure}

\begin{table} 
\caption {Real lenses statistical data in RMSp for short and long runs.}

\centering
  \begin{tabular}{|c|c|c|c|c|}
    \hline
   \textbf{Algorithm}&	\textbf{Mean}& \textbf{Std. dev.}& \textbf{Min.}& \textbf{Max.}\\ \hline    
       \multicolumn{5}{|c|}{\textbf{Short runs 1 million}} \\ \hline
     {\bf S-Newton} & 5.3910E+00       & 3.7429E+00 & 1.7935E+00       & 1.5780E+01 \\ \hline
     {\bf IF 10}    & 1.8657E+01       & 1.1726E+01 & 2.0361E+00       & 3.9179E+01 \\ \hline
     {\bf IF 20}    & 5.4893E+00       & 2.9982E+00 & 1.4134E+00       & 1.0765E+01 \\ \hline
     {\bf IF 50}    & {\bf 5.2665E+00} & 3.1108E+00 & {\bf 1.1120E+00} & 1.0768E+01 \\ \hline
     {\bf IF 100}   & {\bf 4.8658E+00} & 2.6844E+00 & {\bf 1.1081E+00} & 1.0781E+01 \\ \hline
       \multicolumn{5}{|c|}{\textbf{Long runs 4 million}} \\ \hline
       {\bf L-Newton} & {\bf 4.6116E+00} & 2.6540E+00 & {\bf 1.1254E+00} & 1.0327E+01 \\ \hline
       {\bf IF 40}    & 1.8218E+01       & 1.1024E+01 & 6.3836E+00       & 3.9179E+01 \\ \hline
       {\bf IF 80}    & 5.5141E+00       & 2.9581E+00 & 1.7265E+00       & 1.0764E+01 \\ \hline
       {\bf IF 200}   & 5.2529E+00       & 2.9357E+00 & 1.9582E+00       & 1.0766E+01 \\ \hline
       {\bf FI 400}   & {\bf 4.9059E+00} & 2.6582E+00 & {\bf 1.3586E+00} & 1.0786E+01 \\ \hline
  \end{tabular}
\label{stat02}       

\end{table}

\begin{figure} 
\centering
\includegraphics[width=15cm]{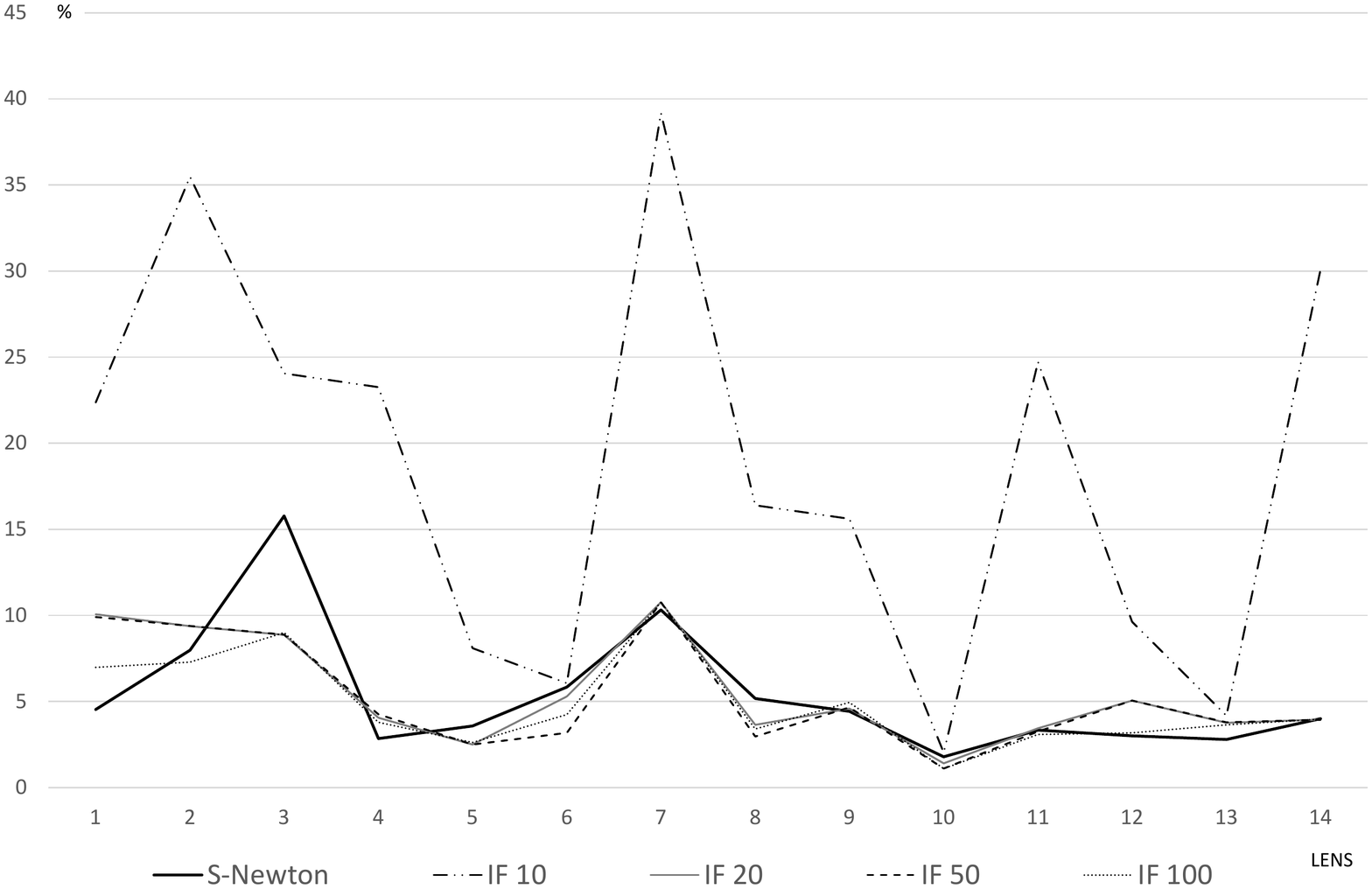}
\caption{Best found solution on a short run. Real lenses per algorithm.}
\label{shortreal}
\end{figure}

\begin{figure} 
\centering
\includegraphics[width=15cm]{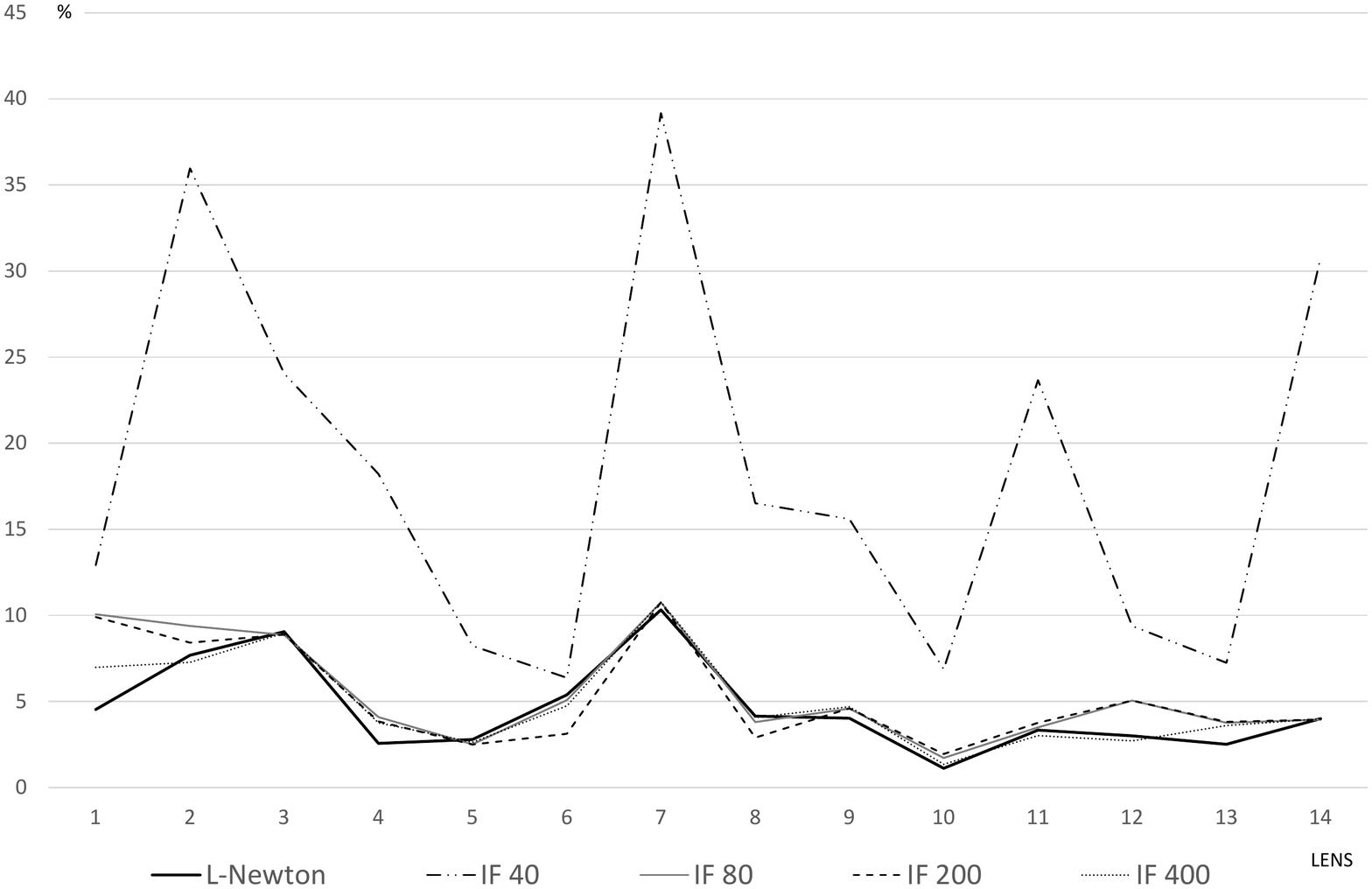}
\caption{Best found solution on a long run. Real lenses per algorithm.}
\label{longreal}
\end{figure}

Experimetal results are given in Figures \ref{shortart}, \ref{longart}, \ref{shortreal},   \ref{longreal}   
and  are summarized in Table  \ref{stat01} and  Table   \ref{stat02}.

\medskip\noindent
{\bf Comparison of the algorithms based on raw experimental results.}
Obviously, the pure multi-start Newton method  is by far the best on artificial instances.

On the real lenses, the situation is a bit different. The Iterative improvement on several occasions outperforms the 
Newton method  with random initial solutions.
On both datasets, the long run yields only slightly better results as the short run does  
(and  the short run is four times faster in executing).
While we have no idea how far from optimal solutions the achieved values are for real lenses, 
we know that, by construction, a solution with 0\% RMS error exists for each of the artificial lenses. Because of that it is worth to note that on the artificial set, the random algorithms found nearly optimal solutions in all cases. The RMS errors are in the range of E-04, which still is not pure 0\% RMS error, but the very small difference could be due to rounding of the values in the .ies files.

On the other hand we did not find very low RMS values on the real set. The values that were found corresponded with the values of previous test that were performed without any numerical assistance. We did however perform an experiment on the real set with longer running time, in which we generated 16 million and 64 million initial solutions that showed similar behaviour. 
The mean error and the minimum error over 14 lenses decreased under 3.0\% and 1.50\% with 16 million generated solutions, and under 2.0\% and 1.0\% RMS error after 64 million.

\begin{figure}[h]
\centering
\includegraphics[width=15cm]{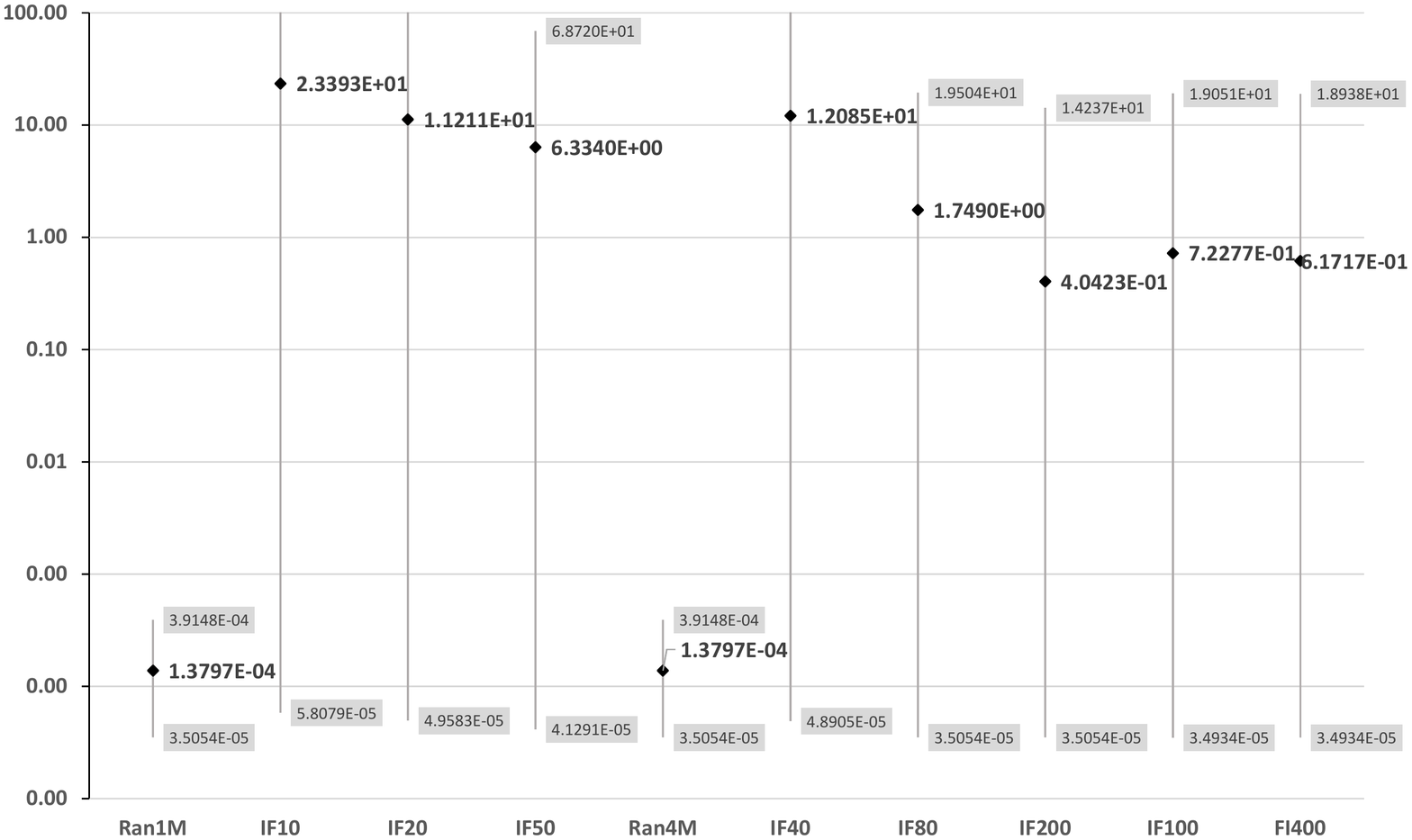}
\caption{Min-Max scatter diagram for artificial lenses. Logarithmic y axis!}
\label{scatter}
\end{figure}

While the success of the Newton method on artificial lenses is not surprising, it  is
not clear  why the method is struggling on the realistic dataset.  
The winners in this comparison are on both sets the same, but   on the real set the differences between the random Newton and IF assisted Newton algorithms are a lot smaller than on the artificial set, we even see that on some instances the IF algorithms are better. This could be due to the fact that IF algorithm
was previously developed for real lenses that are from a limited range in search space and thus 
has a slight advantage on the set. 
The advantage of the pure Newton method on the artificial dataset can nicely be observed  from  
 the data scatter in Figure \ref{scatter}. 
We see that the IF algorithms provide very high degree of data scatter where the random ones provide very narrow result window. This may be due to the nature of the search because the IF algorithms focus in one defined direction which may not be the best one. Because of that the Newton optimization can not escape the potential pitfall of the direction. 
In contrast, the randomly generated solutions generally find lower quality results that at the same time provide more 
maneuvering space for the Newton method to find the best direction on more takes.                        
 
\medskip\noindent
{\bf A  comparison of the algorithms based on weighted ranking.}

We assign  a weight from 1 to 10 to each instance solution per algorithm.
If the algorithm found the best solution on an instance it would get the weight 10 and if it found the second best solution it would get the weight 9, and so on until 1 for the worst solution. 
The total score of the algorithm is the sum of the scores on each instance.

In the same way, we compute the score  based on the average values per algorithm and lens. The results are presented in Table \ref{rank}.     
\begin{table}[]
\scriptsize
\centering
\caption{Weight based ranking.}
\label{my-label}
\begin{tabular}{|l|l|r|l|r|l|r|l|r|}
\hline
\multirow{4}{*}{\begin{tabular}[c]{@{}l@{}}\textbf{R}\\ \textbf{A}\\ \textbf{N}\\ \textbf{K}\end{tabular}} & \multicolumn{4}{l|}{} & \multicolumn{4}{l|}{}                                                                                                                                                                                                                                                            \\ 
                                                                       & \multicolumn{4}{c|}{{\bf Artificial}}                                                                                                   & \multicolumn{4}{c|}{{\bf Real}}                                                                                                         \\ \cline{2-9} 
{\bf }                                                                 & \multicolumn{2}{c|}{{\bf Best}}                                    & \multicolumn{2}{c|}{{\bf Mean}}                                    & \multicolumn{2}{c|}{{\bf Best}}                                    & \multicolumn{2}{c|}{{\bf Mean}}                                    \\ \cline{2-9}
{\bf }                                                                 & \multicolumn{1}{c|}{{\bf Alg.}} & \multicolumn{1}{c|}{{\bf Score}} & \multicolumn{1}{c|}{{\bf Alg.}} & \multicolumn{1}{c|}{{\bf Score}} & \multicolumn{1}{c|}{{\bf Alg.}} & \multicolumn{1}{c|}{{\bf Score}} & \multicolumn{1}{c|}{{\bf Alg.}} & \multicolumn{1}{c|}{{\bf Score}} \\ \hline
1                                                                      & L-Newton                        & 944                              & L-Newton                        & 1000                             & L-Newton                        & 102                              & L-Newton                        & 140                              \\ \hline
2                                                                      & S-Newton                        & 940                              & S-Newton                        & 900                              & IF 100                          & 97                               & S-Newton                        & 124                              \\ \hline
3                                                                      & IF 200                          & 836                              & IF 20                           & 588                              & IF 400                          & 96                               & IF 10                           & 114                              \\ \hline
4                                                                      & IF 80                           & 810                              & IF 100                          & 585                              & IF 50                           & 94                               & IF 50                           & 89                               \\ \hline
5                                                                      & IF 100                          & 678                              & IF 10                           & 559                              & S-Newton                        & 90                               & IF 20                           & 77                               \\ \hline
6                                                                      & IF 400                          & 660                              & IF 50                           & 538                              & IF 20                           & 88                               & IF 100                          & 72                               \\ \hline
7                                                                      & IF 50                           & 529                              & IF 400                          & 385                              & IF 200                          & 86                               & IF 40                           & 61                               \\ \hline
8                                                                      & IF 40                           & 514                              & IF 40                           & 358                              & IF 80                           & 83                               & IF 80                           & 39                               \\ \hline
9                                                                      & IF 20                           & 379                              & IF 80                           & 304                              & IF 10                           & 23                               & IF 400                          & 34                               \\ \hline
10                                                                     & IF 10                           & 212                              & IF 200                          & 283                              & IF 40                           & 21                               & IF 200                          & 20                               \\ \hline
\end{tabular}
\label{rank} 
\end{table}


Note that the ranking here compares both short and long runs. 
As expected,  Table \ref{rank}  confirms the superiority of the pure Newton method on the artificial dataset.
However,  the situation is much more complicated on the dataset of real lenses. Despite the Newton method (long and short run) being the two best when considering the average results, they are not both when looking at the best solutions! The long run Newton is still the overall best, but the short run Newton is on the fifth place outran by two short IF algorithms and one long IF algorithm. We can also observe that the score differences are much smaller on the real set, which indicates that the pure Newton method is not as superior as it was on the artificial set.  The lesser superiority could be explained in part by fact that the IF algorithms were developed using the real lens set. Hence the IF could have some unexpected advantages. However the Newton method improves the results in all cases.

\medskip\noindent
{\bf Wilcoxon test.}
The third comparison is based on  the statistical paired singed Wilcoxon \cite{Wilcoxon} rank test. The statistical test compares algorithms pair by pair to estimate the difference between them. This is done via the asymptotic difference. If the value of the asymptotic difference is lower than $0.05$ then the algorithms in pair significantly differ one from another. The asymptotic differences of algorithm pairs will be presented in tables \ref{st1},\ref{st2},\ref{st3},  and \ref{st4}. 
\begin{table}
 \fontsize{8pt}{12pt}\selectfont
 \caption {Asymptotic significances of Wilcoxon Signed rank test for results for short runs on artificial lenses.}
 \centering
   \begin{tabular}{|l|c|c|c|c|c|}
    \hline
    \textbf{ALG} &  \textbf{IF 20}     & \textbf{IF 50} & \textbf{IF 100}     & \textbf{S-Newton}     \\ \hline
    \textbf{IF 10}              & 4.078E-11 & 1.020E-13      & 1.000E-13 & 1.000E-13 \\ \hline
    \textbf{IF 20}               &                & 7.162E-04      & 3.120E-13 & 1.000E-13 \\ \hline
    \textbf{IF 50}              &                &            & 3.628E-06 & 1.650E-13 \\ \hline
    \textbf{IF 100}               &                &            &                & 3.234E-08 \\ \hline
    \end{tabular}
  \label{st1}
 \end{table}              

\begin{table} 
 \fontsize{8pt}{12pt}\selectfont
 \caption {Asymptotic significances of Wilcoxon Signed rank test for results for long runs on artificial lenses.}
 \centering
   \begin{tabular}{|l|c|c|c|c|c|}
    \hline
    \textbf{ALG} &  \textbf{IF 80}     & \textbf{IF 200} & \textbf{IF 400}     & \textbf{L-Newton}     \\ \hline
    \textbf{IF 40}               & 1.736E-11 & 3.917E-11      & 7.950E-09 & 1.899E-12 \\ \hline
    \textbf{IF 80}               &                & 3.269E-11      & \textbf{5.563E-01} & 3.411E-06 \\ \hline
    \textbf{IF 200}               &                &            & 4.818E-06 & 7.044E-05 \\ \hline
    \textbf{IF 400}               &                &            &                & 1.176E-08 \\ \hline
    \end{tabular}
  \label{st2}
 \end{table}  

\begin{table} 
 \fontsize{8pt}{12pt}\selectfont
 \caption {Asymptotic significances of Wilcoxon Signed rank test for results for short runs on real lenses.}
 \centering
   \begin{tabular}{|l|c|c|c|c|c|}
    \hline
    \textbf{ALG}  & \textbf{IF 20}     & \textbf{IF 50} & \textbf{IF 100}     & \textbf{S-Newton}     \\ \hline
    \textbf{IF 10}             & 9.815E-04 & 9.815E-04      & 9.815E-04 & 9.815E-04 \\ \hline
    \textbf{IF 20}          &                & \textbf{4.326E-01}      & 4.133E-02 & \textbf{5.936E-01} \\ \hline
    \textbf{IF 50}               &                &            & \textbf{5.098E-01} & \textbf{9.750E-01} \\ \hline
    \textbf{IF 100}            &                &            &                & \textbf{4.703E-01} \\ \hline
    \end{tabular}
  \label{st3}
 \end{table}              

\begin{table} 
 \fontsize{8pt}{12pt}\selectfont
 \caption {Asymptotic significances of Wilcoxon Signed rank test for results for long runs on real lenses.}
 \centering
   \begin{tabular}{|l|c|c|c|c|c|}
    \hline
    \textbf{ALG} & \textbf{IF 80}     & \textbf{IF 200} & \textbf{IF 400}     & \textbf{L-Newton}     \\ \hline
    \textbf{IF 40}               & 9.815E-04 & 9.815E-04      & 9.815E-04 & 9.815E-04 \\ \hline
    \textbf{IF 80}               &                & \textbf{5.509E-01}     & \textbf{5.553E-02} & 3.546E-02 \\ \hline
    \textbf{IF 200}             &                &            & \textbf{4.703E-01} & \textbf{1.240E-01} \\ \hline
    \textbf{IF 400}              &                &            &                & \textbf{4.703E-01} \\ \hline
    \end{tabular}
  \label{st4}
 \end{table}  

A look over the Wilcoxon test results reveals that there mostly are no similarities between algorithms when they ran on the artificial set. We can see that the Asymptotic significance values are very low, which means that there are significant differences between algorithms in pairs. We do however have one exception that is the pair IF 80 - IF 400, where the asymptotic difference is just over the margin, so we could say that these two have some similarities. The story is completely different on the real 
dataset, where we can find that most IF algorithms are similar to random algorithms. This also corresponds with the findings of the ranking and RMS error comparison. The Wilcoxon test provided similar conclusions as the previous tests did, but we need to be careful because the data sets differ in size and the real lenses set can be a bit inconclusive as it is a rather small sample with only 14 instances. That is why the artificial lenses with 100 instances could give a more accurate result. 
\begin{table}[]
\centering
\caption{. Real lens RMSp for RAN 4M with and without Newton’s method. Quality increase $\Delta$ in \%.}
\label{my-label}
\begin{tabular}{|l|l|l|l|}
\hline
\textbf{Instance} & \textbf{RAN 4M} & \textbf{Newton} & \textbf{$\Delta$} \\ \hline
CP12632           & 27,996          & 7,6908          & 72,53      \\ \hline
CP12634           & 45,8986         & 9,05513         & 80,27      \\ \hline
CP12633           & 10,7706         & 2,57185         & 76,12      \\ \hline
CA11934           & 10,2492         & 2,7982          & 72,70      \\ \hline
CA11268           & 15,1818         & 5,38851         & 64,51      \\ \hline
CP12817           & 29,6252         & 10,3279         & 65,14      \\ \hline
CA11265           & 9,64366         & 4,1553          & 56,91      \\ \hline
CP12636           & 7,62895         & 4,03813         & 47,07      \\ \hline
CA13013           & 2,70647         & 1,12548         & 58,42      \\ \hline
FP13030           & 10,1866         & 3,34882         & 67,13      \\ \hline
CA11525           & 12,0224         & 3,00557         & 75,00      \\ \hline
CA12392           & 6,87747         & 2,51916         & 63,37      \\ \hline
CA11483           & 24,5813         & 4,0032          & 83,71      \\ \hline
\end{tabular}
\end{table}

\section{Conclusion}
 
Here we presented an upgrade of a previously developed most promising discrete optimization heuristics with a numerical approach to optimization. It was shown that application of  Newton’s method gave improvement of both performance and quality of solutions. In terms of raw performance, we got from the initial 8 minutes’ runtime for one algorithm on one lens to about 45s runtime with the upgraded IF 10 or S-Newton algorithm. The stated runtime is accurate for symmetric lenses and an input of 91 vectors. When working on asymmetric lenses the input will be around 33.000 vectors, this is when the problem becomes a big data problem. But because of the algorithms construction the runtime should increase to about 2 hours and 15 minutes. The increase will be by a factor of 180, while the amount of vectors is increased by a factor 360. On the asymmetric lenses the runtime will be lowered from around 24 hours to 2 hours and 15 minutes. Despite the drastic time shortening the quality of the solutions was not worse thanks to Newton’s method, which enabled us to find local minimums on the majority of solutions found by the heuristic algorithms. Newton’s method in fact successfully minimized the RMS error on all of the experiment cases with the average of 60\% increased quality (minimized RMS) over previous experiments done in \cite{OpticsExpress, informatica}. This can be well observed in Table 9, where we can see the RMS error found by the RAN 4M algorithm before the application of Newton’s method and after. We can conclude that the integration of numerical approach with previously developed heuristics significantly improved the application performance to the level at which it is useful in the main research. On the other hand, we have learned that due to sensitiveness of  Newton’s method to the choice of initial solutions, it may be rewarding to use a preprocessor that may provide promising initial solutions. In particular, on the dataset consisting of real lenses, the experiment showed that the initial solutions provided by a discrete local search algorithm improved the overall performance of the algorithm. This leads to the conclusion that a combination of an algorithm that finds promising initial solutions as a preprocessor to Newton’s method may be a winning combination, at least on some datasets of instances. Hence, in a practical application, it may be worth developing good heuristics that may handle specific properties of the instances and thus provide promising initial solutions for final optimization.

\section{Acknowledgement}
This work was supported in part by ARRS, the Research agency of Slovenia, grants P1-0285 
 and ARRS-1000-15-0510. We sincerely thank both reviewers for carefully reading the manuscript and for constructive remarks.
 
\section*{References}

\bibliography{newton_bib}

\appendix
\section{}
\footnotesize

\textbf{Evaluation function :}
\begin{scriptsize}
\begin{equation}
 E \left(\textbf{a},\textbf{b},\textbf{c}\right) 
= \frac{1}{N}\sum_{i=1}^N \left[I_{max} ( a_{1}\cos^{c_{1}}(\varPhi_{i}-b_{1}) +a_{2}\cos^{c_{2}}(\varPhi_{i}-b_{2}) +a_{3}\cos^{c_{3}}(\varPhi_{i}-b_{3}) ) -  I_m(\varPhi_i)  \right] ^{2}  
\label{eval1}
\end{equation}
\end{scriptsize}

\textbf{Partial equations:}

\begin{equation}
G_{i} = I_{max} ( a_{1}\cos^{c_{1}}(\varPhi_{i}-b_{1}) +a_{2}\cos^{c_{2}}(\varPhi_{i}-b_{2}) +a_{3}\cos^{c_{3}}(\varPhi_{i}-b_{3}) ) -  I_m(\varPhi_i)
\label{g}
\end{equation}

\begin{equation}
F_{1_{i}} = I_{max} \cos^{c_{1}}(\varPhi_{i}-b_{1})
\label{f1}
\end{equation}

\begin{equation}
F_{2_{i}} = I_{max} \cos^{c_{2}}(\varPhi_{i}-b_{2})
\label{f2}
\end{equation}

\begin{equation}
F_{3_{i}} = I_{max} \cos^{c_{3}}(\varPhi_{i}-b_{3})
\label{f3}
\end{equation}

\begin{equation}
F_{4_{i}} = I_{max} a_{1}c_{1}\cos^{(c_{1}-1)}(\varPhi_{i}-b_{1})\sin(\varPhi_{i}-b_{1})
\label{f4}
\end{equation}

\begin{equation}
F_{5_{i}} = I_{max} a_{2}c_{2}\cos^{(c_{2}-1)}(\varPhi_{i}-b_{2})\sin(\varPhi_{i}-b_{2})
\label{f5}
\end{equation}

\begin{equation}
F_{6_{i}} = I_{max} a_{3}c_{3}\cos^{(c_{3}-1)}(\varPhi_{i}-b_{3})\sin(\varPhi_{i}-b_{3})
\label{f6}
\end{equation}

\begin{equation}
F_{7_{i}} = I_{max} a_{1}\cos^{c_{1}}(\varPhi_{i}-b_{1})\ln(\cos(\varPhi_{i}-b_{1}))
\label{f7}
\end{equation}

\begin{equation}
F_{8_{i}} = I_{max} a_{2}\cos^{c_{2}}(\varPhi_{i}-b_{2})\ln(\cos(\varPhi_{i}-b_{2}))
\label{f8}
\end{equation}

\begin{equation}
F_{9_{i}} = I_{max} a_{3}\cos^{c_{3}}(\varPhi_{i}-b_{3})\ln(\cos(\varPhi_{i}-b_{3}))
\label{f9}
\end{equation}

\begin{equation}
S_{14_{i}} = I_{max} c_{1}\cos^{(c_{1}-1)}(\varPhi_{i}-b_{1})\sin(\varPhi_{i}-b_{1})
\label{s14}
\end{equation}

\begin{equation}
S_{17_{i}} = I_{max} \cos^{c_{1}}(\varPhi_{i}-b_{1})\ln(\cos(\varPhi_{i}-b_{1}))
\label{s17}
\end{equation}

\begin{equation}
S_{25_{i}} = I_{max} c_{2}\cos^{(c_{2}-1)}(\varPhi_{i}-b_{2})\sin(\varPhi_{i}-b_{2})
\label{s25}
\end{equation}

\begin{equation}
S_{28_{i}} = I_{max} \cos^{c_{2}}(\varPhi_{i}-b_{2})\ln(\cos(\varPhi_{i}-b_{2}))
\label{s28}
\end{equation}

\begin{equation}
S_{36_{i}} = I_{max} c_{3}\cos^{(c_{3}-1)}(\varPhi_{i}-b_{3})\sin(\varPhi_{i}-b_{3})
\label{s36}
\end{equation}

\begin{equation}
S_{39_{i}} = I_{max} \cos^{c_{3}}(\varPhi_{i}-b_{3})\ln(\cos(\varPhi_{i}-b_{3}))
\label{s39}
\end{equation}

\begin{equation}
S_{41_{i}} = I_{max} c_{1}\cos^{(c_{1}-1)}(\varPhi_{i}-b_{1})\sin(\varPhi_{i}-b_{1}))
\label{s41}
\end{equation}

\begin{footnotesize}
\begin{equation}
S_{44_{i}} = I_{max} a_{1} c_{1} \cos^{(c_{1}-2)}(\varPhi_{i}-b_{1})(c_{1}\sin^{2}(\varPhi_{i}-b_{1})-1)
\label{s44}
\end{equation}
\end{footnotesize}

\begin{equation}
S_{47_{i}} = -I_{max} a_{1} \cos^{(c_{1}-1)}(\varPhi_{i}-b_{1}) \sin(\varPhi_{i}-b_{1})(c_{1}\ln(\cos(\varPhi_{i}-b_{1}))+1)  
\label{s47}
\end{equation}

\begin{equation}
S_{52_{i}} = I_{max} c_{2}\cos^{(c_{2}-1)}(\varPhi_{i}-b_{2})\sin(\varPhi_{i}-b_{2}))
\label{s52}
\end{equation}

\begin{footnotesize}
\begin{equation}
S_{55_{i}} = I_{max} a_{2} c_{2} \cos^{(c_{2}-2)}(\varPhi_{i}-b_{2})(c_{2}\sin^{2}(\varPhi_{i}-b_{2})-1)
\label{s55}
\end{equation}
\end{footnotesize}

\begin{equation}
S_{58_{i}} = -I_{max} a_{2} \cos^{(c_{2}-1)}(\varPhi_{i}-b_{2}) \sin(\varPhi_{i}-b_{2})(c_{2}\ln(\cos(\varPhi_{i}-b_{2}))+1)
\label{s58}
\end{equation}

\begin{equation}
S_{63_{i}} = I_{max} c_{3}\cos^{(c_{3}-1)}(\varPhi_{i}-b_{3})\sin(\varPhi_{i}-b_{3}))
\label{s63}
\end{equation}

\begin{footnotesize}
\begin{equation}
S_{66_{i}} = I_{max} a_{3} c_{3} \cos^{(c_{3}-2)}(\varPhi_{i}-b_{3})(c_{3}\sin^{2}(\varPhi_{i}-b_{3})-1)
\label{s66}
\end{equation}
\end{footnotesize}

\begin{equation}
S_{69_{i}} =  -I_{max} a_{3} \cos^{(c_{3}-1)}(\varPhi_{i}-b_{3}) \sin(\varPhi_{i}-b_{3})(c_{3}\ln(\cos(\varPhi_{i}-b_{3}))+1) 
\label{s69}
\end{equation}

\begin{equation}
S_{71_{i}} = I_{max} \cos^{c_{1}}(\varPhi_{i}-b_{1})\ln(\cos(\varPhi_{i}-b_{1}))
\label{s71}
\end{equation}

\begin{footnotesize}
\begin{equation}
S_{74_{i}} = -I_{max}(a_{1}\cos^{c_{1}-1}(\varPhi_{i}-b_{1})\sin(\varPhi_{i}-b_{1}) - a_{1} c_{1} \ln(\cos(\varPhi_{i}-b_{1})) \cos^{c_{1}-1}(\varPhi_{i}-b_{1})\sin(\varPhi_{i}-b_{1})) 
\label{s74}
\end{equation}
\end{footnotesize}

\begin{equation}
S_{77_{i}} = I_{max} a_{1}\cos^{c_{1}}(\varPhi_{i}-b_{1})\ln(\cos(\varPhi_{i}-b_{1}))\ln(\cos(\varPhi_{i}-b_{1}))
\label{s77}
\end{equation}

\begin{equation}
S_{82_{i}} = I_{max} \cos^{c_{2}}(\varPhi_{i}-b_{2})\ln(\cos(\varPhi_{i}-b_{2}))
\label{s82}
\end{equation}

\begin{footnotesize}
\begin{equation}
S_{85_{i}} = -I_{max}(a_{2}\cos^{c_{2}-1}(\varPhi_{i}-b_{2})\sin(\varPhi_{i}-b_{2}) - a_{2} c_{2} \ln(\cos(\varPhi_{i}-b_{2})) \cos^{c_{2}-1}(\varPhi_{i}-b_{2})\sin(\varPhi_{i}-b_{2}))
\label{s85}
\end{equation}
\end{footnotesize}

\begin{equation}
S_{88_{i}} = I_{max} a_{2}\cos^{c_{2}}(\varPhi_{i}-b_{2})\ln(\cos(\varPhi_{i}-b_{2}))\ln(\cos(\varPhi_{i}-b_{2}))
\label{s88}
\end{equation}

\begin{equation}
S_{93_{i}} = I_{max} \cos^{c_{3}}(\varPhi_{i}-b_{3})\ln(\cos(\varPhi_{i}-b_{3}))
\label{s93}
\end{equation}

\begin{footnotesize}
\begin{equation}
S_{96_{i}} = -I_{max}(a_{3}\cos^{c_{3}-1}(\varPhi_{i}-b_{3})\sin(\varPhi_{i}-b_{3}) - a_{3} c_{3} \ln(\cos(\varPhi_{i}-b_{3})) \cos^{c_{3}-1}(\varPhi_{i}-b_{3})\sin(\varPhi_{i}-b_{3}))
\label{s96}
\end{equation}
\end{footnotesize}

\begin{equation}
S_{99_{i}} = I_{max} a_{3}\cos^{c_{3}}(\varPhi_{i}-b_{3})\ln(\cos(\varPhi_{i}-b_{3}))\ln(\cos(\varPhi_{i}-b_{3}))
\label{s99}
\end{equation}

\textbf{First order derivatives:}

\begin{equation*}
\frac{\partial E (\textbf{a},\textbf{b},\textbf{c})}{\partial a_{1}} = \sum_{i=1}^{N}G_{i}F_{1_{i}}
\qquad
\frac{\partial E (\textbf{a},\textbf{b},\textbf{c})}{\partial a_{2}} = \sum_{i=1}^{N}G_{i}F_{2_{i}}
\end{equation*}
\begin{equation*}
\frac{\partial E (\textbf{a},\textbf{b},\textbf{c})}{\partial a_{3}} = \sum_{i=1}^{N}G_{i}F_{3_{i}}
\qquad
\frac{\partial E (\textbf{a},\textbf{b},\textbf{c})}{\partial b_{1}} = \sum_{i=1}^{N}G_{i}F_{4_{i}}
\end{equation*}
\begin{equation}
\frac{\partial E (\textbf{a},\textbf{b},\textbf{c})}{\partial b_{2}} = \sum_{i=1}^{N}G_{i}F_{5_{i}}
\qquad
\frac{\partial E (\textbf{a},\textbf{b},\textbf{c})}{\partial b_{3}} = \sum_{i=1}^{N}G_{i}F_{6_{i}}
\label{firstder}
\end{equation}
\begin{equation*}
\frac{\partial E (\textbf{a},\textbf{b},\textbf{c})}{\partial c_{1}} = \sum_{i=1}^{N}G_{i}F_{7_{i}}
\qquad
\frac{\partial E (\textbf{a},\textbf{b},\textbf{c})}{\partial c_{2}} = \sum_{i=1}^{N}G_{i}F_{8_{i}}
\end{equation*}
\begin{equation*}
\frac{\partial E (\textbf{a},\textbf{b},\textbf{c})}{\partial c_{3}} = \sum_{i=1}^{N}G_{i}F_{9_{i}}
\end{equation*}

\textbf{Second order derivatives:}
\begin{equation*}
\frac{\partial^{2} E (\textbf{a},\textbf{b},\textbf{c})}{\partial a_{1}^{2}} = \sum_{i=1}^{N}F_{1_{i}}F_{1_{i}}
\qquad
\frac{\partial^{2} E (\textbf{a},\textbf{b},\textbf{c})}{\partial a_{2}\partial a_{1}} = \sum_{i=1}^{N}F_{1_{i}}F_{2_{i}}
\end{equation*}
\begin{equation*}
\frac{\partial^{2} E (\textbf{a},\textbf{b},\textbf{c})}{\partial a_{3}\partial a_{1}} = \sum_{i=1}^{N}F_{1_{i}}F_{3_{i}}
\qquad
\frac{\partial^{2} E (\textbf{a},\textbf{b},\textbf{c})}{\partial b_{1}\partial a_{1}} = \sum_{i=1}^{N}F_{1_{i}}F_{4_{i}}+G_{i}S_{14_{i}}
\end{equation*}
\begin{equation*}
\frac{\partial^{2} E (\textbf{a},\textbf{b},\textbf{c})}{\partial b_{2}\partial a_{1}} = \sum_{i=1}^{N}F_{1_{i}}F_{5_{i}}
\qquad
\frac{\partial^{2} E (\textbf{a},\textbf{b},\textbf{c})}{\partial b_{3}\partial a_{1}} = \sum_{i=1}^{N}F_{1_{i}}F_{6_{i}}
\end{equation*}
\begin{equation*}
\frac{\partial^{2} E (\textbf{a},\textbf{b},\textbf{c})}{\partial c_{1}\partial a_{1}} = \sum_{i=1}^{N}F_{1_{i}}F_{7_{i}}+G_{i}S_{17_{i}}
\qquad
\frac{\partial^{2} E (\textbf{a},\textbf{b},\textbf{c})}{\partial c_{2}\partial a_{1}} = \sum_{i=1}^{N}F_{1_{i}}F_{8_{i}}
\end{equation*}
\begin{equation*}
\frac{\partial^{2} E (\textbf{a},\textbf{b},\textbf{c})}{\partial c_{3}\partial a_{1}} = \sum_{i=1}^{N}F_{1_{i}}F_{9_{i}}
\end{equation*}

\begin{equation*}
\frac{\partial^{2} E (\textbf{a},\textbf{b},\textbf{c})}{\partial a_{1}\partial a_{2}} = \sum_{i=1}^{N}F_{2_{i}}F_{1_{i}}
\qquad
\frac{\partial^{2} E (\textbf{a},\textbf{b},\textbf{c})}{\partial a_{2}^{2}} = \sum_{i=1}^{N}F_{2_{i}}F_{2_{i}}
\end{equation*}
\begin{equation*}
\frac{\partial^{2} E (\textbf{a},\textbf{b},\textbf{c})}{\partial a_{3}\partial a_{2}} = \sum_{i=1}^{N}F_{2_{i}}F_{3_{i}}
\qquad
\frac{\partial^{2} E (\textbf{a},\textbf{b},\textbf{c})}{\partial b_{1}\partial a_{2}} = \sum_{i=1}^{N}F_{2_{i}}F_{4_{i}}
\end{equation*}
\begin{equation*}
\frac{\partial^{2} E (\textbf{a},\textbf{b},\textbf{c})}{\partial b_{2}\partial a_{2}} = \sum_{i=1}^{N}F_{2_{i}}F_{5_{i}}+G_{i}S_{25_{i}}
\qquad
\frac{\partial^{2} E (\textbf{a},\textbf{b},\textbf{c})}{\partial b_{3}\partial a_{2}} = \sum_{i=1}^{N}F_{2_{i}}F_{6_{i}}
\end{equation*}
\begin{equation*}
\frac{\partial^{2} E (\textbf{a},\textbf{b},\textbf{c})}{\partial c_{1}\partial a_{2}} = \sum_{i=1}^{N}F_{2_{i}}F_{7_{i}}
\qquad
\frac{\partial^{2} E (\textbf{a},\textbf{b},\textbf{c})}{\partial c_{2}\partial a_{2}} = \sum_{i=1}^{N}F_{2_{i}}F_{8_{i}}+G_{i}S_{28_{i}}
\end{equation*}
\begin{equation*}
\frac{\partial^{2} E (\textbf{a},\textbf{b},\textbf{c})}{\partial c_{3}\partial a_{2}} = \sum_{i=1}^{N}F_{2_{i}}F_{9_{i}}
\end{equation*}

\begin{equation*}
\frac{\partial^{2} E (\textbf{a},\textbf{b},\textbf{c})}{\partial a_{1}\partial a_{3}} = \sum_{i=1}^{N} F_{3_{i}}F_{1_{i}}
\qquad
\frac{\partial^{2} E (\textbf{a},\textbf{b},\textbf{c})}{\partial a_{2}\partial a_{3}} = \sum_{i=1}^{N}F_{3_{i}}F_{2_{i}}
\end{equation*}
\begin{equation*}
\frac{\partial^{2} E (\textbf{a},\textbf{b},\textbf{c})}{\partial a_{3}^{2}} = \sum_{i=1}^{N}F_{3_{i}}F_{3_{i}}
\qquad
\frac{\partial^{2} E (\textbf{a},\textbf{b},\textbf{c})}{\partial b_{1}\partial a_{3}} = \sum_{i=1}^{N}F_{3_{i}}F_{4_{i}}
\end{equation*}
\begin{equation*}
\frac{\partial^{2} E (\textbf{a},\textbf{b},\textbf{c})}{\partial b_{2}\partial a_{3}} = \sum_{i=1}^{N}F_{3_{i}}F_{5_{i}}
\qquad
\frac{\partial^{2} E (\textbf{a},\textbf{b},\textbf{c})}{\partial b_{3}\partial a_{3}} = \sum_{i=1}^{N}F_{3_{i}}F_{6_{i}}+G_{i}S_{35_{i}}
\end{equation*}
\begin{equation*}
\frac{\partial^{2} E (\textbf{a},\textbf{b},\textbf{c})}{\partial c_{1}\partial a_{3}} = \sum_{i=1}^{N}F_{3_{i}}F_{7_{i}}
\qquad
\frac{\partial^{2} E (\textbf{a},\textbf{b},\textbf{c})}{\partial c_{2}\partial a_{3}} = \sum_{i=1}^{N}F_{3_{i}}F_{8_{i}}
\end{equation*}
\begin{equation*}
\frac{\partial^{2} E (\textbf{a},\textbf{b},\textbf{c})}{\partial c_{3}\partial a_{3}} = \sum_{i=1}^{N}F_{3_{i}}F_{9_{i}}+G_{i}S_{39_{i}}
\end{equation*}

\begin{equation*}
\frac{\partial^{2} E (\textbf{a},\textbf{b},\textbf{c})}{\partial a_{1}\partial b_{1}} = \sum_{i=1}^{N}F_{4_{i}}F_{1_{i}}+GS_{41_{i}}
\qquad
\frac{\partial^{2} E (\textbf{a},\textbf{b},\textbf{c})}{\partial a_{2}\partial b_{1}} = \sum_{i=1}^{N}F_{4_{i}}F_{2_{i}}
\end{equation*}
\begin{equation*}
\frac{\partial^{2} E (\textbf{a},\textbf{b},\textbf{c})}{\partial a_{3}\partial b_{1}} = \sum_{i=1}^{N}F_{4_{i}}F_{3_{i}}
\qquad
\frac{\partial^{2} E (\textbf{a},\textbf{b},\textbf{c})}{\partial b_{1}ˇ{2}} = \sum_{i=1}^{N}F_{4_{i}}F_{4_{i}}+G_{i}S_{44_{i}}
\end{equation*}
\begin{equation*}
\frac{\partial^{2} E (\textbf{a},\textbf{b},\textbf{c})}{\partial b_{2}\partial b_{1}} = \sum_{i=1}^{N}F_{4_{i}}F_{5_{i}}
\qquad
\frac{\partial^{2} E (\textbf{a},\textbf{b},\textbf{c})}{\partial b_{3}\partial b_{1}} = \sum_{i=1}^{N}F_{4_{i}}F_{6_{i}}
\end{equation*}
\begin{equation*}
\frac{\partial^{2} E (\textbf{a},\textbf{b},\textbf{c})}{\partial c_{1}\partial b_{1}} = \sum_{i=1}^{N}F_{4_{i}}F_{7_{i}}+G_{i}S_{47_{i}}
\qquad
\frac{\partial^{2} E (\textbf{a},\textbf{b},\textbf{c})}{\partial c_{2}\partial b_{1}} = \sum_{i=1}^{N}F_{4_{i}}F_{8_{i}}
\end{equation*}
\begin{equation*}
\frac{\partial^{2} E (\textbf{a},\textbf{b},\textbf{c})}{\partial c_{3}\partial b_{1}} = \sum_{i=1}^{N}F_{4_{i}}F_{9_{i}}
\end{equation*}

\begin{equation*}
\frac{\partial^{2} E (\textbf{a},\textbf{b},\textbf{c})}{\partial a_{1}\partial b_{2}} = \sum_{i=1}^{N}F_{5_{i}}F_{1_{i}}
\qquad
\frac{\partial^{2} E (\textbf{a},\textbf{b},\textbf{c})}{\partial a_{2}\partial b_{2}} = \sum_{i=1}^{N}F_{5_{i}}F_{2_{i}}+G_{i}S_{52_{i}}
\end{equation*}
\begin{equation*}
\frac{\partial^{2} E (\textbf{a},\textbf{b},\textbf{c})}{\partial a_{3}\partial b_{2}} = \sum_{i=1}^{N}F_{5_{i}}F_{3_{i}}
\qquad
\frac{\partial^{2} E (\textbf{a},\textbf{b},\textbf{c})}{\partial b_{1}\partial b_{2}} = \sum_{i=1}^{N}F_{5_{i}}F_{4_{i}}
\end{equation*}
\begin{equation*}
\frac{\partial^{2} E (\textbf{a},\textbf{b},\textbf{c})}{\partial b_{2}^{2}} = \sum_{i=1}^{N}F_{5_{i}}F_{5_{i}}+G_{i}S_{55_{i}}
\qquad
\frac{\partial^{2} E (\textbf{a},\textbf{b},\textbf{c})}{\partial b_{3}\partial b_{2}} = \sum_{i=1}^{N}F_{5_{i}}F_{6_{i}}
\end{equation*}
\begin{equation*}
\frac{\partial^{2} E (\textbf{a},\textbf{b},\textbf{c})}{\partial c_{1}\partial b_{2}} = \sum_{i=1}^{N}F_{5_{i}}F_{7_{i}}
\qquad
\frac{\partial^{2} E (\textbf{a},\textbf{b},\textbf{c})}{\partial c_{2}\partial b_{2}} = \sum_{i=1}^{N}F_{5_{i}}F_{8_{i}}+G_{i}S_{58_{i}}
\end{equation*}
\begin{equation*}
\frac{\partial^{2} E (\textbf{a},\textbf{b},\textbf{c})}{\partial c_{3}\partial b_{2}} = \sum_{i=1}^{N}F_{5_{i}}F_{9_{i}}
\end{equation*}

\begin{equation*}
\frac{\partial^{2} E (\textbf{a},\textbf{b},\textbf{c})}{\partial a_{1}\partial b_{3}} = \sum_{i=1}^{N} F_{6_{i}}F_{1_{i}}
\qquad
\frac{\partial^{2} E (\textbf{a},\textbf{b},\textbf{c})}{\partial a_{2}\partial b_{3}} = \sum_{i=1}^{N}F_{6_{i}}F_{2_{i}}
\end{equation*}
\begin{equation*}
\frac{\partial^{2} E (\textbf{a},\textbf{b},\textbf{c})}{\partial a_{3}^{2}} = \sum_{i=1}^{N}F_{6_{i}}F_{3_{i}}+GS_{63_{i}}
\qquad
\frac{\partial^{2} E (\textbf{a},\textbf{b},\textbf{c})}{\partial b_{1}\partial b_{3}} = \sum_{i=1}^{N}F_{6_{i}}F_{4_{i}}
\end{equation*}
\begin{equation*}
\frac{\partial^{2} E (\textbf{a},\textbf{b},\textbf{c})}{\partial b_{2}\partial b_{3}} = \sum_{i=1}^{N}F_{6_{i}}F_{5_{i}}
\qquad
\frac{\partial^{2} E (\textbf{a},\textbf{b},\textbf{c})}{\partial b_{3}^{2}} = \sum_{i=1}^{N}F_{6_{i}}F_{6_{i}}+G_{i}S_{66_{i}}
\end{equation*}
\begin{equation*}
\frac{\partial^{2} E (\textbf{a},\textbf{b},\textbf{c})}{\partial c_{1}\partial b_{3}} = \sum_{i=1}^{N}F_{6_{i}}F_{7_{i}}
\qquad
\frac{\partial^{2} E (\textbf{a},\textbf{b},\textbf{c})}{\partial c_{2}\partial b_{3}} = \sum_{i=1}^{N}F_{6_{i}}F_{8_{i}}
\end{equation*}
\begin{equation*}
\frac{\partial^{2} E (\textbf{a},\textbf{b},\textbf{c})}{\partial c_{3}\partial b_{3}} = \sum_{i=1}^{N}F_{6_{i}}F_{9_{i}}+G_{i}S_{69_{i}}
\end{equation*}

\begin{equation*}
\frac{\partial^{2} E (\textbf{a},\textbf{b},\textbf{c})}{\partial a_{1}\partial c_{1}} = \sum_{i=1}^{N}F_{7}F_{1_{i}}+GS_{71_{i}}
\qquad
\frac{\partial^{2} E (\textbf{a},\textbf{b},\textbf{c})}{\partial a_{2}\partial c_{1}} = \sum_{i=1}^{N}F_{7_{i}}F_{2_{i}}
\end{equation*}
\begin{equation*}
\frac{\partial^{2} E (\textbf{a},\textbf{b},\textbf{c})}{\partial a_{3}\partial c_{1}} = \sum_{i=1}^{N}F_{7_{i}}F_{3_{i}}
\qquad
\frac{\partial^{2} E (\textbf{a},\textbf{b},\textbf{c})}{\partial b_{1}\partial c_{1}} = \sum_{i=1}^{N}F_{7_{i}}F_{4_{i}}+G_{i}S_{74_{i}}
\end{equation*}
\begin{equation*}
\frac{\partial^{2} E (\textbf{a},\textbf{b},\textbf{c})}{\partial b_{2}\partial c_{1}} = \sum_{i=1}^{N}F_{7_{i}}F_{5_{i}}
\qquad
\frac{\partial^{2} E (\textbf{a},\textbf{b},\textbf{c})}{\partial b_{3}\partial c_{1}} = \sum_{i=1}^{N}F_{7_{i}}F_{6_{i}}
\end{equation*}
\begin{equation*}
\frac{\partial^{2} E (\textbf{a},\textbf{b},\textbf{c})}{\partial c_{1}^{2}} = \sum_{i=1}^{N}F_{7_{i}}F_{7_{i}}+G_{i}S_{77_{i}}
\qquad
\frac{\partial^{2} E (\textbf{a},\textbf{b},\textbf{c})}{\partial c_{2}\partial c_{1}} = \sum_{i=1}^{N}F_{7_{i}}F_{8_{i}}
\end{equation*}
\begin{equation*}
\frac{\partial^{2} E (\textbf{a},\textbf{b},\textbf{c})}{\partial c_{3}\partial c_{1}} = \sum_{i=1}^{N}F_{7_{i}}F_{9_{i}}
\end{equation*}

\begin{equation*}
\frac{\partial^{2} E (\textbf{a},\textbf{b},\textbf{c})}{\partial a_{1}\partial c_{2}} = \sum_{i=1}^{N}F_{8_{i}}F_{1_{i}}
\qquad
\frac{\partial^{2} E (\textbf{a},\textbf{b},\textbf{c})}{\partial a_{2}\partial c_{2}} = \sum_{i=1}^{N}F_{8_{i}}F_{2_{i}}+G_{i}S_{82_{i}}
\end{equation*}
\begin{equation*}
\frac{\partial^{2} E (\textbf{a},\textbf{b},\textbf{c})}{\partial a_{3}\partial c_{2}} = \sum_{i=1}^{N}F_{8_{i}}F_{3_{i}}
\qquad
\frac{\partial^{2} E (\textbf{a},\textbf{b},\textbf{c})}{\partial b_{1}\partial c_{2}} = \sum_{i=1}^{N}F_{8_{i}}F_{4_{i}}
\end{equation*}
\begin{equation*}
\frac{\partial^{2} E (\textbf{a},\textbf{b},\textbf{c})}{\partial b_{2}\partial c_{2}} = \sum_{i=1}^{N}F_{8_{i}}F_{5_{i}}+G_{i}S_{85_{i}}
\qquad
\frac{\partial^{2} E (\textbf{a},\textbf{b},\textbf{c})}{\partial b_{3}\partial c_{2}} = \sum_{i=1}^{N}F_{8_{i}}F_{6_{i}}
\end{equation*}
\begin{equation*}
\frac{\partial^{2} E (\textbf{a},\textbf{b},\textbf{c})}{\partial c_{1}\partial c_{2}} = \sum_{i=1}^{N}F_{8_{i}}F_{7_{i}}
\qquad
\frac{\partial^{2} E (\textbf{a},\textbf{b},\textbf{c})}{\partial c_{2}^{2}} = \sum_{i=1}^{N}F_{8_{i}}F_{8_{i}}+G_{i}S_{88_{i}}
\end{equation*}
\begin{equation*}
\frac{\partial^{2} E (\textbf{a},\textbf{b},\textbf{c})}{\partial c_{3}\partial c_{2}} = \sum_{i=1}^{N}F_{8_{i}}F_{9_{i}}
\end{equation*}

\begin{equation*}
\frac{\partial^{2} E (\textbf{a},\textbf{b},\textbf{c})}{\partial a_{1}\partial c_{3}} = \sum_{i=1}^{N} F_{9_{i}}F_{1_{i}}
\qquad
\frac{\partial^{2} E (\textbf{a},\textbf{b},\textbf{c})}{\partial a_{2}\partial c_{3}} = \sum_{i=1}^{N}F_{9_{i}}F_{2_{i}}
\end{equation*}
\begin{equation*}
\frac{\partial^{2} E (\textbf{a},\textbf{b},\textbf{c})}{\partial a_{3}^{2}} = \sum_{i=1}^{N}F_{9_{i}}F_{3_{i}}+G_{i}S_{93_{i}}
\qquad
\frac{\partial^{2} E (\textbf{a},\textbf{b},\textbf{c})}{\partial b_{1}\partial c_{3}} = \sum_{i=1}^{N}F_{9_{i}}F_{4_{i}}
\end{equation*}
\begin{equation*}
\frac{\partial^{2} E (\textbf{a},\textbf{b},\textbf{c})}{\partial b_{2}\partial c_{3}} = \sum_{i=1}^{N}F_{9_{i}}F_{5_{i}}
\qquad
\frac{\partial^{2} E (\textbf{a},\textbf{b},\textbf{c})}{\partial b_{3}\partial c_{3}} = \sum_{i=1}^{N}F_{9_{i}}F_{6_{i}}+G_{i}S_{96_{i}}
\end{equation*}
\begin{equation*}
\frac{\partial^{2} E (\textbf{a},\textbf{b},\textbf{c})}{\partial c_{1}\partial c_{3}} = \sum_{i=1}^{N}F_{9_{i}}F_{7_{i}}
\qquad
\frac{\partial^{2} E (\textbf{a},\textbf{b},\textbf{c})}{\partial c_{2}\partial c_{3}} = \sum_{i=1}^{N}F_{9_{i}}F_{8_{i}}
\end{equation*}
\begin{equation*}
\frac{\partial^{2} E (\textbf{a},\textbf{b},\textbf{c})}{\partial c_{3}^{2}} = \sum_{i=1}^{N}F_{9_{i}}F_{9_{i}}+G_{i}S_{99_{i}}
\end{equation*}

\textbf{Jacobian matrix:}
\[
J(a_{1}, ... , c_{3}) =
\begin{bmatrix} 

\dfrac{\partial^{2} E (\textbf{a},\textbf{b},\textbf{c})}{\partial a_{1}\partial a_{1}} & \cdots & \dfrac{\partial^{2} E (\textbf{a},\textbf{b},\textbf{c})}{\partial c_{3}\partial a_{1}} \\ \vdots & \ddots & \vdots \\ \dfrac{\partial^{2} E (\textbf{a},\textbf{b},\textbf{c})}{\partial a_{1}\partial c_{3}} & \cdots & \dfrac{\partial^{2} E (\textbf{a},\textbf{b},\textbf{c})}{\partial c_{3}\partial c_{3}} 

\end{bmatrix}
\]

\textbf{Delta vector:}
\[ 
\textbf{d} =
\begin{bmatrix} 
da_{1}&da_{2}&da_{3}&db_{1}&db_{2}&db_{3}&dc_{1}&dc_{2}&dc_{3}
\end{bmatrix}
\]

\textbf{Right side:}
\[ 
R(a_{1}, ... , c_{3}) =
\begin{bmatrix} 
\dfrac{\partial E (\textbf{a},\textbf{b},\textbf{c})}{\partial a_{1}} & \cdots & \dfrac{\partial E (\textbf{a},\textbf{b},\textbf{c})}{\partial c_{3}}
\end{bmatrix}^{T}
\]

\textbf{System of equations to solve for d:}

\begin{equation}
J(\textbf{x}_{i})\times\textbf{d}_{i} = R(\textbf{x}_{i})
\end{equation}

\textbf{Coeficient vector:}
\[ 
\textbf{x} =
\begin{bmatrix} 
a_{1}&a_{2}&a_{3}&b_{1}&b_{2}&b_{3}&c_{1}&c_{2}&c_{3}
\end{bmatrix}
\]

\textbf{Iterative scheme :}

\begin{equation}
\textbf{x}_{i+1} = \textbf{x}_{i}  -  \textbf{d}_{i} 
\end{equation}

\end{document}